\documentclass[runningheads]{llncs}

\usepackage[T1]{fontenc}
\usepackage{color}
\usepackage{multirow} 
\usepackage{booktabs} 
\usepackage{amsmath}
\usepackage{subfigure}  

\usepackage{caption}
\usepackage{subcaption}
\usepackage{natbib} 

\usepackage{enumitem}
\setlist[itemize]{label=\textbullet}

\usepackage{graphicx}
\newcommand{\pers}{PERS}
\newcommand{\ers}{ERS}

\usepackage{natbib}
\setcitestyle{numbers,square}

\begin{document}
\title{Personalized Programming Guidance based on Deep
Programming Learning Style Capturing}
\titlerunning{Personalized Programming Guidance}
%
\author{Yingfan Liu\inst{1} \and
Renyu Zhu\inst{1,3} \and
Ming Gao\inst{1,2}\thanks{Corresponding author}}
\authorrunning{Yingfan Liu et al.}
%
\institute{School of Data Science \& Engineering \\ 
\and KLATASDS-MOE in School of Statistics \\ 
East China Normal University, Shanghai, China \\
\email{yfliu.ds@stu.ecnu.edu.cn} \\
\email{mgao@dase.ecnu.edu.cn} \\
\and NetEase Fuxi AI Lab, Hangzhou, Zhejiang, China \\
\email{zhurenyu@corp.netease.com} \\
}
\maketitle              
\pagestyle{empty}  
\thispagestyle{empty} 

\begin{abstract}
With the rapid development of big data and AI technology, programming is in high demand and has become an essential skill for students. 
Meanwhile, researchers also focus on boosting the online judging system's guidance ability to reduce students' dropout rates.
%
Previous studies mainly targeted at enhancing learner engagement on online platforms by providing personalized recommendations. 
However, 
two significant challenges still need to be addressed in programming: \textbf{\textit{C1}}) how to recognize complex programming behaviors; \textbf{\textit{C2}}) how to capture intrinsic learning patterns that align with the actual learning process.
%
To fill these gaps, 
in this paper, 
we propose a novel model called \textbf{P}rogramming \textbf{E}xercise \textbf{R}ecommender with Learning \textbf{S}tyle (\textbf{PERS}), which simulates learners' intricate programming behaviors. 
%
Specifically, since programming is an iterative and trial-and-error process, we first introduce a positional encoding and a differentiating module to capture the changes of consecutive code submissions (which addresses \textbf{\textit{C1}}). 
To better profile programming behaviors, 
we extend the Felder-Silverman learning style model~\footnote{A theory proposed by Richard M. Felder and Linda K. Silverman to describe individuals' preferred ways of learning. Further details will be provided in the related work section.}, a classical pedagogical theory, to perceive intrinsic programming patterns. 
Based on this, we align three latent vectors to record and update programming ability, processing style, and understanding style, respectively (which addresses \textbf{\textit{C2}}). 
We perform extensive experiments on two real-world datasets to verify the rationality of modeling programming learning styles and the effectiveness of PERS for personalized programming guidance.

\end{abstract}
\section{Introduction}
The rapid advancement of AI technology has profoundly influenced on individuals of diverse backgrounds and skill levels. In this connection, online judge systems have emerged as an indispensable avenue for those seeking to boost their programming proficiency. 
However, despite the growing popularity of this learning modality, high dropout rates have been observed, attributable to the inadequate provision of personalized instructions tailored to learners' unique learning preferences~\cite{DBLP:conf/aied/PereiraOCFSAAA19}.

%


Recently, recommender systems have been widely applied in online education scenarios to facilitate personalized learning. 
There are various recommendation models, including CF(collaborative filtering)-based methods~\cite{ DBLP:conf/cikm/MaoZWDDXH21,DBLP:conf/sigir/WuWF0CLX21}, content-based methods~\cite{DBLP:journals/eswa/BagherHM17, DBLP:conf/icsai/HuangL18} and deep-learning-based methods~\cite{DBLP:journals/corr/abs-2203-11011, DBLP:conf/sigir/Gong0WFP0Y20}. 
These general models aim to provide personalized recommendations by capturing users' interests and needs through static preferences and individual interactions. 
%
%
In the context of programming, however, the learning process exhibits a dynamic and progressive nature. 
%
%
This represents an essential application of the sequential recommendation (SR) task, which predicts subsequent behavioral sequences based on historical records~\cite{DBLP:conf/sigir/HouHZZ22,  DBLP:conf/cikm/SunLWPLOJ19, DBLP:conf/recsys/TanXL16, DBLP:conf/aaai/WuT0WXT19}. 
While extant SR models have yielded successful results in e-learning contexts, there remain significant gaps in directly deploying them to programming scenarios~\cite{DBLP:conf/sigir/LiYDSLSC22}. 
As illustrated by Figure~\ref{fig:plsc_data_model}, programming learning differs from traditional learning in two crucial respects: i) it enables learners to make multiple attempts on the same exercise and edit their previous submissions based on the feedback received from the compiler and, ii) the platform can record fine-grained behavioral data related to programming, including code snippets, compilation time, and compilation status.
%
Furthermore, current sequential models prioritize learners' patterns with little regard to their intrinsic behaviors, including learning styles. 
These styles reflect the ways in which learners process and comprehend information, and are thus factors that cannot be ignored.

Consequently, the study of SR in programming learning confronts two significant challenges. 
First, it is imperative to model distinctive and fine-grained patterns involved in programming, including code-related side features and iterative submission behavior(\textbf{\textit{C1}}). 
Second, there is an urgent need to incorporate pedagogical theory into the model to bolster its interpretability with the actual learning process(\textbf{\textit{C2})}.

%

To address the above challenges, we propose a new model named \textbf{P}rogramming \textbf{E}xercise \textbf{R}ecommender with Learning \textbf{S}tyle (PERS). 
To simulate the iterative process in programming, we employ a two-step approach. 
Firstly, we map programming exercises and code-related features (such as code snippets, execution time, and execution status) into embeddings using a representation module with positional encoding. 
Secondly, we formulate a differentiating module that calculates the changes between consecutive code submissions. 
This module can adeptly capture fine-grained learning patterns by effectively distinguishing between intra-exercise or inter-exercise attempts (for \textbf{\textit{C1}}).
To enhance the consistency between our proposed model and the actual learning process, we draw inspiration from a pedagogical theory known as the Felder-Silverman Learning Style Model (FSLSM)~\cite{felder1988learning}, which is widely utilized in educational scenarios for mining learning patterns and delivering personalized guidance.
%
Considering  the processing and understanding dimensions in FSLSM, we present a formal definition and detailed descriptions of programming learning styles in this paper. 
On this foundation, we develop three latent vectors: programming ability, processing style, and understanding style, which are designed to track the learners' intrinsic behavioral patterns during the programming process (for \textbf{\textit{C2}}). 
After obtaining the above vectors, our model employs a multilayer perceptron (MLP) to generate personalized predictions that align individuals' learning preferences.
%
%
In summary, the main contributions of this paper are summarized as follows:
\begin{itemize}
    \item Our study endeavors to furnish personalized programming guidance by emulating the iterative and trial-and-error programming learning process, thereby offering a novel vantage point on programming education. 
    \item We have meaningfully incorporated the FSLSM pedagogical theory into our model, enabling us to effectively capturing the intrinsic behavioral patterns of students while also enhancing rationality and consistency.
    \item We conduct experiments on two real-world datasets to validate the efficacy and interpretability of our approach.
\end{itemize}


\section{Related Works}

\subsection{Sequential Recommendation}
Sequential recommendation models aim to incorporate users' personalized and contextual information based on their historical interactions~\cite{DBLP:conf/icdm/KangM18} to predict the future behaviors. \par
In earlier studies, researchers considered Markov chains as a powerful method to model interaction processes and capture users' sequential patterns~\cite{DBLP:conf/icdm/HeM16, DBLP:conf/ijcai/CaiHM17}. Later, the advent of recurrent neural networks (RNN) greatly has expanded the potential of recommender systems to process multi-type input data and understand complex item transitions. For example, ~\cite{DBLP:journals/corr/HidasiKBT15} first adopt RNN on real-life session-based recommendations and then enhance the backbone with parallel RNN to leverage more rich features~\cite{DBLP:conf/recsys/HidasiQKT16}.
There are various techniques designed to improve the RNN-based models, such as data augmentation (GRU4Rec~\cite{DBLP:conf/recsys/TanXL16}), data reconstruction (SRGNN~\cite{DBLP:conf/aaai/WuT0WXT19}), and unified module injection (SINE~\cite{DBLP:conf/wsdm/TanZYLZYH21}, CORE~\cite{DBLP:conf/sigir/HouHZZ22}). Recently, another line of works has seeked to use the Transformer module to capture global information, which RNN overlooks. For instance, BERT4Rec~\cite{DBLP:conf/cikm/SunLWPLOJ19} utilize bidirectional self-attention with Cloze tasks during training to enhance the hidden representation. \par




\subsection{Sequential Recommendation in E-learning}
%
Existing research on SR in e-learning typically focus on recommending the most appropriate resources, such as courses and exercises, to learners by capturing their static and dynamic characteristics through their past behavioral record~\cite{DBLP:conf/sdm/0007WC0Y0F21}. 
%
For instance, ~\cite{DBLP:conf/adma/JiangZWLY22} and ~\cite{DBLP:conf/cscwd/MaHTZ22} propose a cognitive diagnostic method to model students’ proficiency on each exercises based on probabilistic matrix factorization and students’ proficiency. ~\cite{DBLP:journals/kbs/RenLSZ23} apply a knowledge tracing model with an enhanced self-attention to measures students’ mastery states and assist model to recommend. These methods effectively capture students' preferences and mastery of knowledge points. However, they often overlook the impact of students' internal learning styles.

In the field of programming, some preliminary attempts have been made to explore the personalized recommendations.
For example, ~\cite{DBLP:conf/sigir/LiYDSLSC22} apply BERT~\cite{DBLP:conf/naacl/DevlinCLT19} to encode students' source code and propose a knowledge tracing model to capture mastery of programming skills. 
%
However, the dynamic sequential patterns in existing works are not consistent with real programming process due to ignore the iterative process. 
%


\subsection{Learning Style Model}
%
Learning styles refer to the way in which students prefer to obtain, process and retain information~\cite{felder2005applications}. The most common theoretical models include Felder-Silverman Learning Style Model (FSLSM), Kolb's learing style~\cite{kolb1999learning} and VARK model~\cite{prithishkumar2014understanding}.
Previous research has demonstrated that the FSLSM is more comprehensible and appropriate for identifying learning styles in online learning compared to other models~\cite{DBLP:journals/eswa/MuhammadQWA22}. 
This model describes learning styles from four dimensions: the perspective of perception (sensitive/intuitive), information input (visual/verbal), processing (active/reflective) and understanding (sequential/global) based on the learner’s behavior patterns during learning process. \par

\section{Preliminaries}

\subsection{Programming Learning Style Model}
%
%
Inspired by the FSLSM, we define a programming learning style model (PLSM) centered around the problem-solving behavior observed in online judging systems. 

As shown in Table~\ref{tab:chap3_behavior_example}, the PLSM delineates the inherent learning patterns during programming through two dimensions: processing and understanding. 
In terms of processing, learners can be classified as either active or reflective. 
%
When solving exercises, active learners tend to think through a complete answer before submitting their solution, while reflective learners prefer to attempt the same exercise multiple times and refine their previous submissions based on the compiler feedback. 
As for the dimension of understanding, learners can be labeled as sequential or global. 
Sequential learners tend to approach learning tasks in a progressive sequence, such as in numerical or knowledge concept order. 
In contrast, 
global learners tend to approach tasks in a non-linear fashion,
such as by selecting tasks that they find most interesting or engaging.
These distinct learning styles reflect learners' preferences and can significantly impact the trajectory of their problem-solving process.


\begin{table}
\caption{Programming Learning Style Model}
  \begin{tabular}{lcp{7.5cm}}
    \toprule
    \textbf{Dimension} & \textbf{Label} & \textbf{Learning Characteristics}\\ 
    \midrule
    \multirow{2}*{Processing} & Active & Solving exercises by figuring out a complete answer before submitting.\\ 
    & Reflective & Solving exercises through multiple trials and errors.  \\ 
    \midrule 
    \multirow{2}*{Understanding} & Sequential & Solving step-by-step \\  & Global & Solving by leaps and bounds \\ 
    \bottomrule
  \end{tabular}
  
  \label{tab:chap3_behavior_example}
\end{table}




\subsection{Problem Definition}

\begin{figure}[htbp]
    \centering
    \includegraphics[width=1\textwidth]{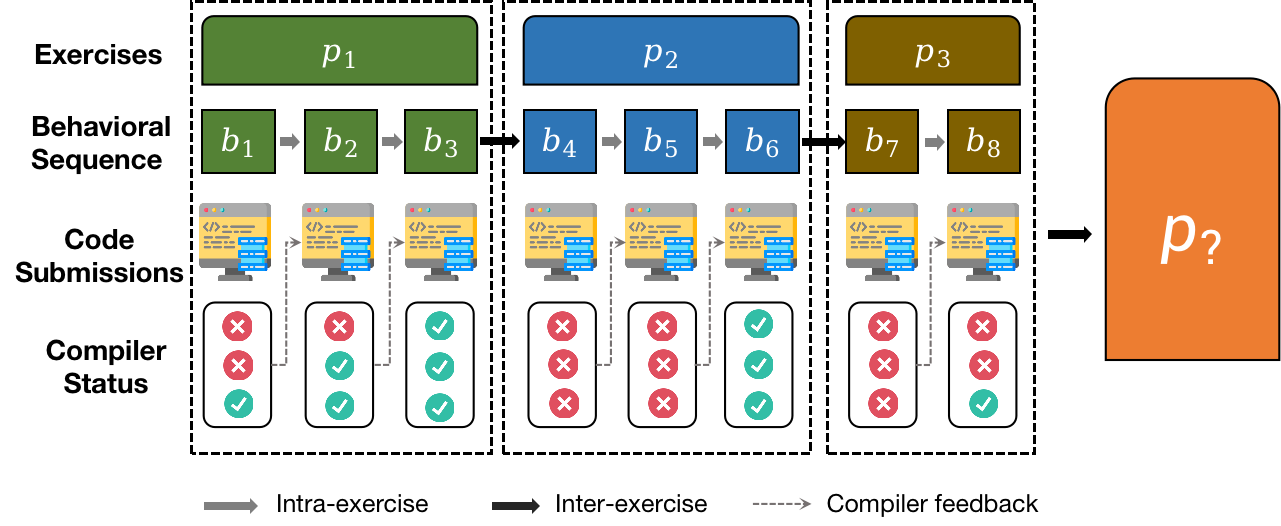}
    \caption{Data model for PER task}
    \label{fig:plsc_data_model}
\end{figure}

To foster learners' involvement and enhance their programming skills in online judge systems, 
we present a new task called programming exercise recommendation (PER). 
The definition of PER is as follows and an example of data model is depicted in Figure~\ref{fig:plsc_data_model}.

\textbf{\textit{Definition: Programming Exercise Recommendation.}}  
Suppose there are $n$ online learning users $\mathcal{U} = \{u_1, u_2, \cdots, u_n\}$ with problem-solving behavior logs $\mathcal{B}= \{B_1, B_2, \cdots, B_n\}$ and $m$ programming exercises $\mathcal{P} = \{p_1, p_2, \cdots, p_m\}$. 
Specifically, the $i$-th record $B_i$ =  $\{b_1, b_2, \cdots, b_{l_i}\}$ represents the interaction sequence of the $i$-th learner $u_i$, 
where $l_i$ represents the length of the sequence.
Each element $b_j$ in the sequence is a triple $\langle p_{b_j}, c_{b_j}, r_{b_j} \rangle$ consisting of the problem $p_{b_j}$, 
the code $c_{b_j}$ and the compilation result $r_{b_j}$.
The ultimate goal of programming exercise recommendation is to predict learners' learning preferences in the future based on the past interaction behavior $B_i$ between learners and exercises. 
that is, the next exercise $p_{b_{li}+1}$ that will be tried. 
Correspondingly, in machine learning methods, the optimization objective is:
\begin{small}
\begin{align}
   l_i = & \max_{\mathcal{A}} \prod_{(b_i, p_{b_{l_i + 1}}) \in \mathcal{B}\cup \mathcal{B}^{-}}\log \mathcal{A}(p_{b_{l_i + 1}}|B_i)^{y_i}(1-\mathcal{A}(p_{b_{l_i + 1}})|B_i)^{(1-y_i)},
\end{align}
\end{small}
where $\mathcal{A}$ is a probabilistic prediction model, 
such as neural networks, whose output is to predict the probability of interacting with the next exercise $p_{b_{li}+1}$ based on the historical behavior sequence $B_i$. 
$\mathcal{B}^{-}$ is a set of negative samples, i.e.,
the exercises that learner $u_i$ has not interacted with label $y_i = 1$ if and only if $(B_i, p_{b_{l_i + 1}}) \in \mathcal{B}$,
otherwise $y_i$ = 0.



\section{PERS Framework}
\begin{figure*}[htbp]
    \includegraphics[width=1\textwidth]{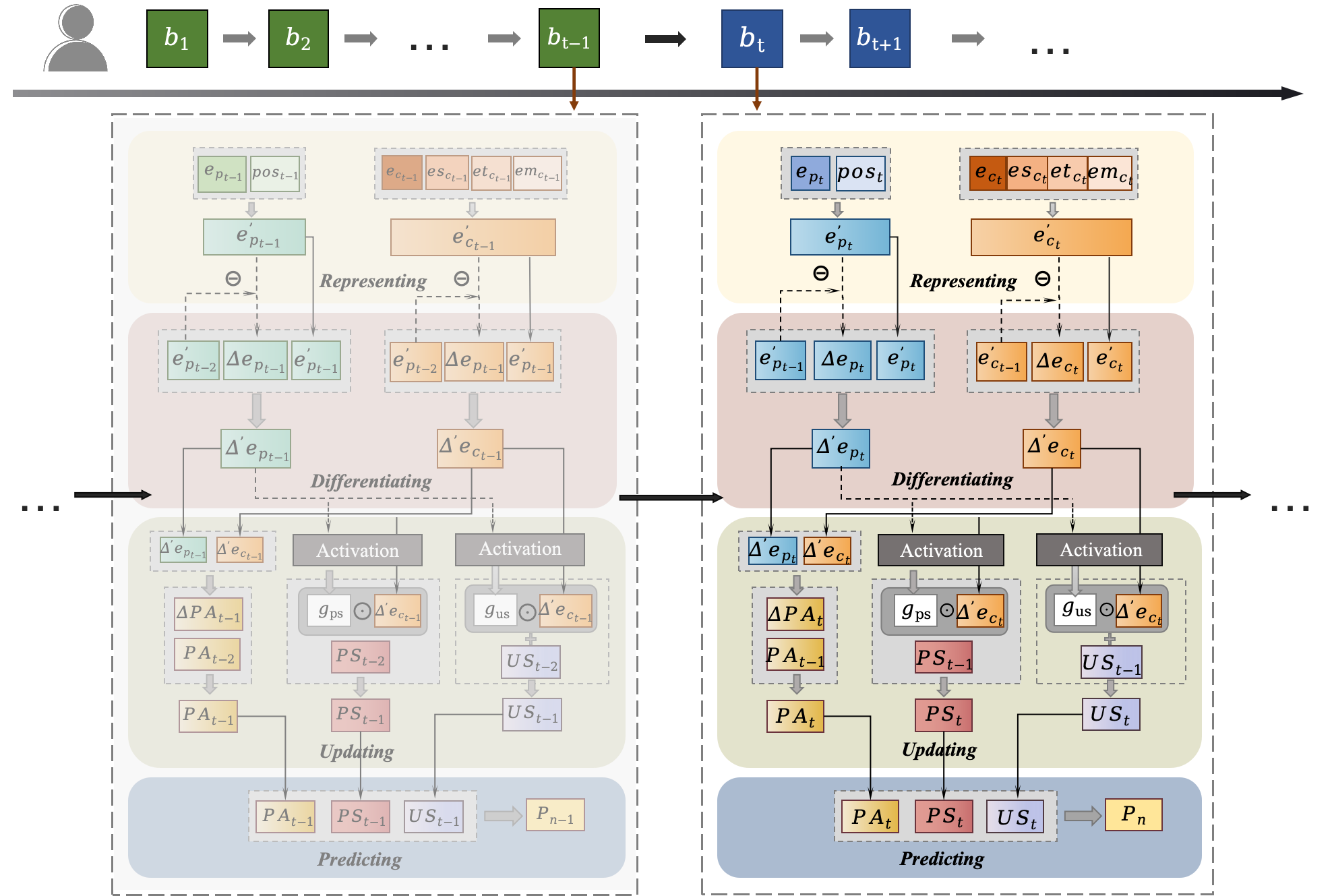}
    \caption{\pers\ Architecture}
    \label{fig:chap4_plsc_arc}
\end{figure*}

In this section, 
we propose a deep learning framework, namely \pers, to solve programming exercise recommendation.
As shown in Figure~\ref{fig:chap4_plsc_arc}, 
the architecture of \pers\ is mainly composed of four functional modules: representing, differentiating, updating and predicting.
%
The details of the four modules are given in the following.

\subsection{Representing Module}
The representing module mainly focuses on obtaining the embedding of the two inputs: 
exercises and codes.

\subsubsection{Exercise Representation}
As demonstrated in Figure~\ref{fig:plsc_data_model},
learners typically attempt a programming exercise multiple times until they pass all test cases. 
Even when trying the same exercise, 
the compilation results for each attempt are distinct and progressive.
Therefore,
the exercise embedding and the positional embedding in the sequence are both critical.
Suppose $p_t$ denotes the programming exercise coded by the learners at time $t$.
First, we use a projection matrix $\mathbf{E}_p \in \mathcal{R}^{(N+2)\times d_p}$ to represent each exercise by its id, 
where $N$ is the total number of exercises and $d_p$ is the dimension. 
The first dimension of the projection matrix here is $N+2$ because two zero pads are added.
Then the representation vector for the problem $p_t$ can be obtained as $\mathbf{e}_{p_t} \in \mathcal{R}^{d_p}$. 
In addition, inspired by the work~\cite{DBLP:conf/nips/VaswaniSPUJGKP17}, we use the sinusoidal function to acquire the position embedding $\mathbf{pos}_{t}$ at time $t$:
\begin{align}
\mathbf{pos}_{(t, 2i)} & = \sin (t/10000^{2i/d_{pos}}), \\
\mathbf{pos}_{(t, 2i+1)} & = \cos (t/10000^{2i/d_{pos}}).
\end{align}
where $d_{pos}$ denotes the dimension. 
Based on the exercise embedding $\mathbf{e}_{p_t}$ and the position embedding $\mathbf{pos}_t$,
we obtain the enhanced exercise embedding $\mathbf{e}_{p_t}^{'}$ through an MLP:
\begin{align}
   \mathbf{e}_{p_t}^{'} & = \mathbf{W}_1^{T}[\mathbf{e}_{p_t} \oplus \mathbf{pos}_{t}] + \mathbf{b}_1,
\end{align}

where $\oplus$ denotes the vector concatenation, $\mathbf{W}_1 \in \mathcal{R}^{(d_p + d_{pos}) \times d_{k}}$ , $\mathbf{b}_1 \in \mathcal{R}^{d_k}$ are learnable parameters. 

\subsubsection{Code Representation}
Suppose $c_t$ denotes the code submitted by the learners at time step $t$. 
First, we apply a code pre-training model CodeBERT~\cite{DBLP:conf/emnlp/FengGTDFGS0LJZ20} to obtain the initial embedding of code $\mathbf{e}_{c_t} \in \mathcal{R}^{d_c}$.
Additionally, we employ different projection matrices to obtain the representation vectors of code-related side features: the execution time $\mathbf{et}_{c_t}\in \mathcal{R}^{d_{ct}}$, 
the execution memory $\mathbf{em}_{c_t}\in \mathcal{R}^{d_{cm}}$, 
and the execution status $\mathbf{es}_{c_t}\in \mathcal{R}^{d_{cs}}$. 
After all the representation vectors are generated, 
we can obtain the enhanced code embedding $\mathbf{e}_{c_t}^{'}$ by an MLP:
\begin{align}
   \mathbf{e}_{c_t}^{'} & = \mathbf{W}_2^{T}[\mathbf{e}_{c_t} \oplus \mathbf{es}_{c_t} \oplus \mathbf{et}_{c_t} \oplus \mathbf{em}_{c_t}] + \mathbf{b}_2,
\end{align}
where $\mathbf{W}_2 \in \mathcal{R}^{(d_c + d_{cs} + d_{ct} + d_{cm}) \times d_{k}}$ is the weight matrix, $\mathbf{b}_2 \in \mathcal{R}^{d_k}$ is the bias term.

\subsection{Differentiating Module}
As the introduction highlights, one of the challenges in PER is to simulate the iterative and trial-and-error process of programming learning. In this paper, we develop a differentiating module to capture fine-grained learning patterns. 
To distinguish whether students are answering the same exercise or starting a new one, 
we first calculate the exercise difference embedding $\Delta \mathbf{e}_{p_t}$ between students' present exercise embedding $\mathbf{e}_{p_t}^{'}$ and previous exercise embedding $ \mathbf{e}_{p_{t-1}}^{'}$ by subtraction. 
Then, 
we feed the above three embeddings into a multi-layer perceptron to output the final exercise difference embedding $\Delta^{'} \mathbf{e}_{p_t}$:
\begin{align}
    \Delta \mathbf{e}_{p_t} & = \mathbf{e}_{p_t}^{'} - \mathbf{e}_{p_{t-1}}^{'} \\
    \Delta^{'} \mathbf{e}_{p_t} & = \mathbf{W}_3^{T}[\Delta \mathbf{e}_{p_t} \oplus \mathbf{e}_{p_t}^{'} \oplus \mathbf{e}_{p_{t-1}}^{'}] + \mathbf{b}_3, 
\end{align}

For the same exercise,
the codes students submit are different at each attempt, 
which can indicate their progress in the trial-and-error process. 
Therefore, we use students' present code embedding $\mathbf{e}_{c_t}^{'}$,
previous code embedding $\mathbf{e}_{c_{t-1}}^{'}$ and the difference between them $\Delta \mathbf{e}_{c_t}$ to obtain the final code difference embedding $\Delta^{'} \mathbf{e}_{c_t}$:
\begin{align}
    \Delta \mathbf{e}_{c_t} & = \mathbf{e}_{c_t}^{'} - \mathbf{e}_{c_{t-1}}^{'}, \\
    \Delta^{'} \mathbf{e}_{c_t} & = \mathbf{W}_4^{T}[\Delta \mathbf{e}_{c_t} \oplus \mathbf{e}_{c_t}^{'} \oplus \mathbf{e}_{c_{t-1}}^{'}] + \mathbf{b}_4, 
\end{align}
where $\mathbf{W}_3, \mathbf{W}_4 \in \mathcal{R}^{3d_k \times d_{k}}$ is the weight matrix, $\mathbf{b}_3, \mathbf{b}_4 \in \mathcal{R}^{d_k}$ is the bias term.

\subsection{Updating Module}
The purpose of this module is to update the latent states that represent the learner's intrinsic learning style. 
Inspired by the classic learning style model FSLSM,
we propose two hidden vectors, processing style $\mathbf{PS}_{t}$ and understanding style $\mathbf{US}_{t}$, to capture the programming learning style of learners. 
In addition, 
motivated by the programming knowledge tracing research~\cite{DBLP:conf/icdm/ZhuZHGLQZ22}, 
we introduce another hidden vector called programming ability $\mathbf{PA}_{t}$ to enhance the modeling of programming behavior. 

First, 
we assume that all learners start with the same programming ability $\mathbf{PA}_{0}$, and their programming ability will gradually improve as they progress through exercises. 
The learners' programming ability $\mathbf{PA}_{t}$ at time step $t$ depends on their performance in completing the current exercises as well as their previous programming ability $\mathbf{PA}_{t-1}$. The corresponding update process is as follows:
\begin{align}
    \Delta \mathbf{PA} & = \mathbf{W}_5^T[\mathbf{e}_{p_t}^{'} \oplus \mathbf{e}_{c_t}^{'}] + \mathbf{b_5}, \\
    \mathbf{PA}_{t} & = \mathbf{W}_6^T [\Delta \mathbf{PA} \oplus \mathbf{PA}_{t-1}] + \mathbf{b_6},
\end{align}
where $\mathbf{W}_5, \mathbf{W}_6 \in \mathcal{R}^{ 2d_k \times d_{k}}$ are weight matrices,  $\mathbf{b}_5, \mathbf{b}_6 \in \mathcal{R}^{d_k}$ are bias terms. 
When $t=0$, $
\mathbf{PA}_{0} \in \mathcal{R}^{d_k}$ is initialized as a vector of all zeros. 

Similarly, the initial processing style $\mathbf{PS}_{0} \in \mathcal{R}^{d_k}$ at time $t=0$ is also initialized as a vector of all zeros. 
As shown in Table~\ref{tab:chap3_behavior_example}, 
the learner's processing style mainly manifests in their continuous trial-and-error behavior on the same exercise. 
Leveraging the difference of exercise $\Delta^{'} \mathbf{e}_{p_t}$ and code $\Delta^{'} \mathbf{e}_{c_t}$ generated from the difference module, 
we introduce a gating mechanism to update the learner's processing style vector $\mathbf{PS}_{t}$. 
We first calculate a selection gate $\mathbf{g}_{ps}$ using $\Delta^{'} \mathbf{e}_{p_t}$,  
which determines whether the current exercise is identical to the previous one. 
Then $\mathbf{g}_{ps}$ is multiplied by $\Delta^{'} \mathbf{e}_{c_t}$ to figure out how much semantic information should be learned from the code.
Finally, 
we concatenate the result with the previous processing style $\mathbf{PS}_{t-1}$ and employ a multi-layer perception to fuse these vectors as follows:  
\begin{align}
    \mathbf{g}_{ps} & = \tanh (\mathbf{W}_7^T \Delta^{'} \mathbf{e}_{p_t} + \mathbf{b}_7), \label{equ:chap4_gate}\\
    \mathbf{PS}_{t} & = \mathbf{W}_8^T[\mathbf{PS}_{t-1} \oplus (\mathbf{g}_{ps} \odot \Delta^{'} \mathbf{e}_{c_t})] +  \mathbf{b}_8, \label{equ:chap4_ps}
\end{align}
where $\tanh$ is the non-linear activation function, $\mathbf{W}_7 \in \mathcal{R}^{ d_k \times d_{k}}$, $\mathbf{W}_8 \in \mathcal {R}^{ 2d_k \times d_{k}}$, $\mathbf{b}_7, \mathbf{b}_8 \in \mathcal{ R}^{d_k}$ are trainable parameters, $\odot$ is the vector element-wise product operation.

Another latent vector is the understanding style $\mathbf{US}_t$, 
which indicates whether learners prefer to learn step-by-step or in leaps and bounds.
It is derived from the learner's historical records. 
Thus, the initial understanding style $\mathbf{US}_{0} \in \mathcal{R}^{d_k}$ is also initialized as a vector of zeros. 
Similar to the processing style, 
we also employ a gating mechanism to determine whether the learner is attempting the same exercise, 
and subsequently update the current understanding style $\mathbf{US}_t$ based on the previous one $\mathbf{US}_{t-1}$:
\begin{align}
     \mathbf{g}_{us} & = \tanh (\mathbf{W}_9^T \Delta^{'} \mathbf{e}_{p_t} + \mathbf{b}_9), \\
     \mathbf{US}_t & = \mathbf{US}_{t-1} + \mathbf{W}_{10}^T (\mathbf{g}_{us} \odot \mathbf{e}_{p_t }^{'})
\end{align}
where $\mathbf{W}_9, \mathbf{W}_{10} \in \mathcal{R}^{ d_k \times d_{k}}$ are weight matrices, $\mathbf{b}_9 \in \mathcal{R}^{ d_k}$ is the bias term.


\subsection{Predicting Module}
After obtaining the learner's programming ability $\mathbf{PA}_t$,
processing style $\mathbf{PS}_t$, 
and understanding style $\mathbf{US}_t$, 
we can predict the next exercise in the predicting module. 
First, the three intrinsic vectors are concatenated and then projected to the output layer using a fully connected network to get $\mathbf{Pre}_t$. After that, we encode $\mathbf {Pre}_t$ into an $m$-dimensional project matrix and obtain the final probability vector $\mathbf{p}_n$ of exercises being recommended at the next step . 
\begin{align}
     \mathbf{Pre}_t &= \mathbf{W}_{11}^T[\mathbf{PL}_t \oplus \mathbf{PS}_t \oplus \mathbf{US}_t] + \mathbf{b}_{ 11}, \\
     \mathbf{p}_n &= \mathbf{W}_{12}^T \mathbf{Pre}_t + \mathbf{b}_{12}
\end{align}
where $\mathbf{W}_{11} \in \mathcal{R}^{ 3d_k \times d_{k}}$ and $\mathbf{W}_{12} \in \mathcal{R}^{d_k \times d_{n}}$ are weight matrices, 
$\mathbf{b}_{11} \in \mathcal{R} ^{d_k}$ and $\mathbf{b}_{12} \in \mathcal{R} ^{d_m}$ is the bias term.

\section{Experiments}
In this section, we aim to evaluate the effectiveness of \pers\ on programming exercise recommendation through empirical evaluation and answer the following research questions:
\begin{itemize}
    \item \textbf{RQ1}: How does \pers\ perform compared with state-of-the-art pedagogical methods and sequential methods on programming exercise recommendation? 
    \item \textbf{RQ2}: What is the impact of different components on the performance of \pers\ ?
    \item \textbf{RQ3}: How do the primary hyperparameters influence the performance of our model?
    \item \textbf{RQ4}: Can the proposed method learn meaningful intrinsic representations of students during programming?
    
    
\end{itemize}


\subsection{Experimental Settings}
\subsubsection{Datasets}
We evaluate our proposed method \pers\ on two real-world datasets: BePKT~\cite{DBLP:conf/icdm/ZhuZHGLQZ22}and CodeNet~\cite{DBLP:conf/nips/Puri0JZDZD0CDTB21}.
The two datasets are both collected from online judging systems, 
including problems, codes and rich contextual information such as problem descriptions and code compilation results.
Due to the millions of behaviors and contextual data in CodeNet, 
memory overflow exists when processing contextual features such as code and problem descriptions. 
Therefore, 
we sample the CodeNet dataset based on sequence length and submit time,
resulting in two smaller-scale datasets:
CodeNet-len and CodeNet-time.
A brief overview of each dataset is listed as follows:

\begin{itemize}  
    \item \textbf{BePKT}: collected from an online judging system\footnote{\url{https://judgefield.shuishan.net.cn/}} targeted at university education, 
    with its users primarily being college students who start learning to program.
    \item \textbf{CodeNet}: collected and processed by IBM researchers from two large-scale online judgment systems AIZU\footnote{\url{https://onlinejudge.u-aizu.ac.jp/home}} and AtCoder\footnote{\url{https://atcoder.jp/}}. 
    The dataset contains hundreds of thousands of programming learners from different domains.
    \item \textbf{CodeNet-len}: a subset of the CodeNet dataset, which only keeps learners' programming behavioral sequences with lengths between $500$ and $600$.
    \item \textbf{CodeNet-time}: a subset from the CodeNet dataset with submission timestamps between March and April 2020.
\end{itemize}

Table~\ref{tab:chap5_dataset_stat} presents detailed statistics for the above datasets. Specifically, the calculation formula of \#Sparsity is as follows:
\begin{align}\label{equ:sparsity}
\text{\#Sparsity} & = 1 - \frac{\#\text{Interactions}}{\#\text{Students} \times \#\text{Exercises}},
\end{align}

\begin{table}[htbp]
    \centering
    \caption{Detailed statistics of all datasets in experiments, where \#Learners denotes the number of learners, \#Interactions denotes the number of interactions, \#Exercises denotes the number of exercises, \#Sparsity denotes the sparsity of the dataset, \#Pass-Rate denotes the proportion of successful submissions in all submissions, and \#APE(short for Avg-Attempts-Per-Exercise) denotes the average number of attempts on the same programming exercise. 
    }
    \begin{tabular}{c|cccccc}
    \toprule
     Dataset & \#Learners & \#Interactions & \#Exercises & \#Sparsity & \#Pass-Rate & \#APE  \\
     \midrule
     BePKT & 907 & 75,993 & 553 & 84.85\% & 32.03\% & 3.18 \\
     CodeNet  & 154,179  & 13,916,868  & 4,049 & 97.77\% &  53.61\% & 2.05 \\
     CodeNet-time & 26,270 & 811,465 & 2,465 & 98.75\% &  53.42\% & 1.89\\
     CodeNet-len & 1,107 & 605,661 & 3,308 & 83.46\% & 56.88\%  & 1.87\\
    \bottomrule
    \end{tabular}
    \label{tab:chap5_dataset_stat}
\end{table}

\subsubsection{Baselines}
We compare PERS with the following $8$ comparable baselines, which can be grouped into two categories: 
\begin{itemize}
    \item \textbf{Pedagogical methods}: \textbf{ACKRec~\cite{DBLP:conf/sigir/Gong0WFP0Y20}} and \textbf{LPKT~\cite{DBLP:conf/kdd/ShenLCHHYS021}} are two representative methods in e-learning recommendation.
    ACKRec constructs a heterogeneous information network to capture entity relationships. LPKT develops a model by simulating students' learning processes. 
    \item \textbf{Sequential methods}: We introduce $6$ state-of-the-art sequential models, which are 1) \textbf{GRU4Rec}~\cite{DBLP:conf/recsys/TanXL16} introduces data augmentation on recurrent neural network to improve model performance.
    2) \textbf{GRU4Recf}~\cite{DBLP:conf/recsys/HidasiQKT16} further integrates a parallel recurrent neural network to simultaneously represent clicks and feature vectors within interactions.
    3) \textbf{BERT4Rec}~\cite{DBLP:conf/cikm/SunLWPLOJ19} introduces a two-way self-attention mechanism based on BERT~\cite{DBLP:conf/naacl/DevlinCLT19}.
    4) \textbf{SRGNN}~\cite{DBLP:conf/aaai/WuT0WXT19} converts user behavior sequences into graph-structured data and introduces a graph neural network to capture the relationship between items. 
    5)  \textbf{SINE}~\cite{DBLP:conf/wsdm/TanZYLZYH21} proposes a sparse interest network to adaptively generate dynamic preference. 
    6) \textbf{CORE}~\cite{DBLP:conf/sigir/HouHZZ22} designs a representation consistency model to pull the vectors into the same space. 
\end{itemize}

Since all the above baselines do not incorporate code as the model input, for a fair comparison, we implement a degraded version of \pers:
\begin{itemize}
    \item \textbf{\ers}: Remove the code feature input and all subsequent related modules in \pers.
\end{itemize}

\subsubsection{Evaluation Metrics}
To fairly compare different models, inspired by the previous \cite{DBLP:conf/sigir/Gong0WFP0Y20}, 
we choose 
the HR@10 (Hit Ratio),
NDCG@10 (Normalized Discounted Cumulative Gain), 
and MRR@10 (Mean Reciprocal Rank) as the evaluation metrics.

\begin{table*}[htbp]
    \centering
    \caption{The overall performance on two full datasets and two sample datasets. 
    OOM refers to out of memory.}
    \resizebox{\linewidth}{!}{
    \begin{tabular}{c|c|cc|cccccc|cc}
    \toprule
    \multirow{2}{*}{\textbf{Datasets}} & \multirow{2}{*}{\textbf{Metrics}} & \multicolumn{2}{c|}{\textbf{Pedegogicals}} & \multicolumn{6}{c|}{\textbf{Sequentials}} & \multicolumn{2}{c}{\textbf{Ours}}\\
    &  & LPKT & ACKRec & GRU4Rec & GRU4Recf & BERT4Rec & SRGNN & SINE & CORE & \ers & \pers\\
    \midrule
    \multirow{3}{*}{BePKT} & HR@10 & 0.8762 & 0.8849 & 0.9172 & 0.9135 & 0.7369 & 0.9074 & 0.6419 & 0.9172 & \underline{0.9256} & \textbf{0.9288} \\
    & MRR@10  & 0.6743 & 0.6838 & 0.7057 & 0.7053 & 0.5324 & 0.6870 & 0.4059 & 0.6923 & \underline{0.7104} & \textbf{0.7153} \\ 
    & NDCG@10 & 0.7128 & 0.7269 & 0.7573 & 0.7560 & 0.5816 & 0.7408 & 0.4623 & 0.7466 & \underline{0.7645} & \textbf{0.7688} \\
    \midrule
    \multirow{3}{*}{CodeNet} & HR@10 & 0.8423 & 0.8372 & \underline{0.8728} & OOM & 0.6715 & 0.8700 & 0.7276 & 0.8566 & \textbf{0.8803} & OOM \\
    & MRR@10 & 0.6372 & 0.6305 & \underline{0.6927} & OOM & 0.4205 & 0.6879 & 0.4643 & 0.5934 & \textbf{0.7012} & OOM  \\ 
    & NDCG@10 & 0.6983 & 0.6847 & \underline{0.7374} & OOM & 0.4811 & 0.7332 & 0.5285 & 0.6581 & \textbf{0.7435} & OOM \\
    \midrule
    \multirow{3}{*}{CodeNet-len} & HR@10 & 0.7539 & 0.7754 & 0.7812& 0.7767 & 0.1157 & 0.7821 & 0.4602 & 0.7865 & \underline{0.7934} & \textbf{0.8010} \\
    & MRR@10 & 0.5816 & 0.5938 & 0.6189 & 0.6143 & 0.0453 & 0.6073 & 0.2411 & 0.5251 & \underline{0.6235} &\textbf{0.6322} \\ 
    & NDCG@10 & 0.6043 & 0.6139 & 0.6590 & 0.6545 & 0.0614 & 0.6504 & 0.2934 & 0.5850 & \underline{0.6631} & \textbf{0.6704} \\
    \midrule
     \multirow{3}{*}{CodeNet-time} & HR@10 & 0.8532 & 0.8425 & 0.8993 & 0.8991 & 0.7517 & 0.8989 & 0.7742 & 0.8962 & \underline{0.9058} & \textbf{0.9167} \\
    & MRR@10 & 0.6931 & 0.6852 & 0.7309 & 0.7316 & 0.6473 & 0.7236 & 0.5458 & 0.6060 & \underline{0.7422} & \textbf{0.7574} \\ 
    & NDCG@10 & 0.7348 & 0.7233 & 0.7731 & 0.7735 & 0.5978 & 0.7675 & 0.6016 & 0.6775 & \underline{0.7841} & \textbf{0.7923} \\
    \bottomrule
    \end{tabular}
    }
    \label{tab:chap4_exp_result_main}
\end{table*}

\subsubsection{Training Details}
For pedagogical methods, we use the original codes released by their authors~\footnote{\url{https://github.com/JockWang/ACKRec}}~\footnote{\url{https://github.com/bigdata-ustc/EduKTM/tree/main/EduKTM}}. 
Additionally, we implement the PERS model and other baseline models using PyTorch and the RecBole library~\footnote{https://recbole.io/}.
We run all the experiments on a server with 64G memory and two NVIDIA Tesla V100 GPUs.
For all models, 
we set the max sequence length to $50$, 
the batch size of the training set to $2048$ and the test set to $4096$,  
and the optimizer to Adam. 
For the PERS model, we set the exercise and code representation embedding dimensions to 128. 
We perform the hyper-parameter tuning for the learning rate $\{0.1, 0.01, 0.001\}$, the layer number $\{1, 2, 3\}$, and 
 the dropout rate $\{0.1, 0.3, 0.5\}$. 
For all methods, 
we fine-tune the hyperparameters to achieve the best performance and run experiments three times to report the average results. 

\subsection{RQ1: Overall Performance}

Table~\ref{tab:chap4_exp_result_main} summarizes the performance results. We evaluate the methods on four datasets under three evaluation metrics. The best results are highlighted in bold and the best baselines are underlined. 
From results in the Table~\ref{tab:chap4_exp_result_main}, we make the following observations:

\begin{itemize}
    \item 
Our proposed models, PERS and ERS, demonstrate state-of-the-art performance on large-scale programming learning datasets. 
For instance, 
in the case of the CodeNet dataset,
our models exhibit a significant improvement of 1.41\% on HR@10, 1.30\% on MRR@10, 1.12\% on NDCG@10 over the best baseline.

\item 
Code features can significantly improve the performance of the model. In the BePKT, CodeNet-len, and CodeNet-time datasets, the PERS model with code features outperforms the ERS model
%
This finding highlights that code-related features contribute to modeling students' programming learning preferences.
\item 
RNN-based sequential models exhibit superior capabilities in capturing learning behaviors. Our PERS and ERS, which extends recurrent neural networks, achieve the best performance. 
Additionally, the GRU4Rec and GRU4Recf models, designed based on recurrent neural networks, outperform all other sequential methods.
This observation suggests that RNNs are particularly adept at capturing sequential programming behaviors.
\end{itemize}

\subsection{RQ2: Ablation Study}

\begin{figure}
     \centering
     \begin{minipage}{0.45\textwidth}
         \begin{subfigure}
         \centering
            \includegraphics[height=3.5cm]{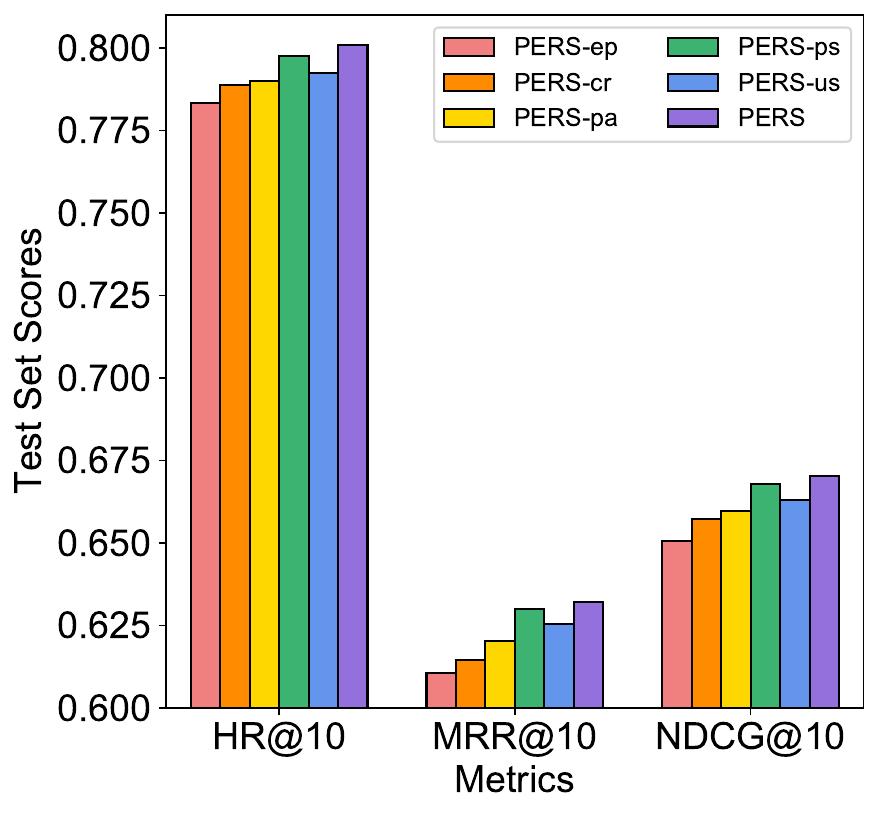}
            \begin{minipage}{\linewidth}
            \centering
              (a) CodeNet-len
            \end{minipage}
            \label{fig:chap4_exp_abl_len} 
         \end{subfigure}
     \end{minipage}
     \begin{minipage}{0.45\textwidth}
          \begin{subfigure}
          \centering
            \includegraphics[height=3.5cm]{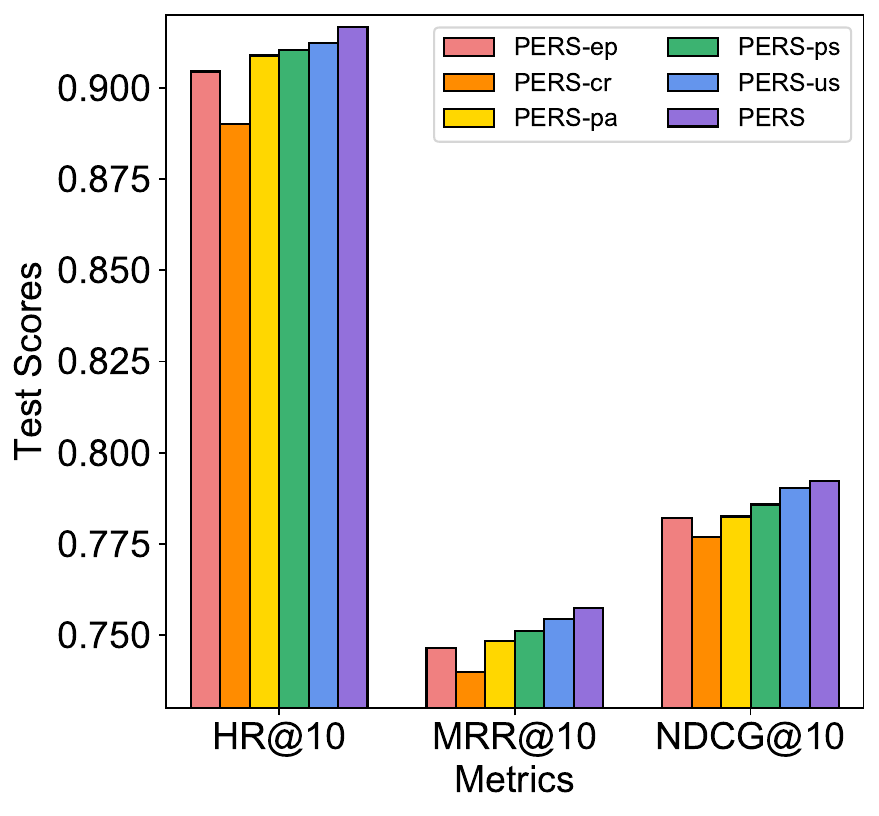}
            \begin{minipage}{\linewidth}
            \centering
              (b) CodeNet-time
            \end{minipage}
            \label{fig:chap4_exp_abl_time}
         \end{subfigure}
     \end{minipage}
     \caption{Ablation Study Results on CodeNet-len(left) and CodeNet-time(right)
     }
    \label{fig:chap4_exp_abl}
\end{figure}

We conduct an ablation study on PERS to understand the importance of the primary components. We obtain five variants: 1) \textbf{\pers-ep}, which removes the exercise position encoding; 2) \textbf{\pers-cr}, which removes the code representation; 3) \textbf{PERS-pa}, 4) \textbf{PERS-ps}, and 5) \textbf{PERS-us} are another three variants that remove $\mathbf{PA_t}$, $\mathbf{PS_t}$ and $\mathbf{US_t}$, respectively. 
Figure~\ref{fig:chap4_exp_abl} displays the results of the PERS model and the its variants on the CodeNet-len and CodeNet-time datasets. From the figure, we can observe: \par

\begin{itemize}
    \item 
Both the representating and updating modules play a crucial role in capturing programming behaviour. As can be observed, the removal of any component of the PERS adversely affects its performance, which emphasizes the rationality and effectiveness of the proposed methods. \par
    \item 
The impact of different components varies across different stages of learning. 
Specifically, in the CodeNet-len dataset, the performance is significantly affected when removing the position encoding of exercises (PERS-ep variant). On the other hand, in the CodeNet-time dataset, the performance sharply declines when the code representation (PERS-cr variant) is removed.
%
This is because the CodeNet-len dataset comprises the latter part of students' behavioral sequences, where students have developed a fixed behavioral pattern. 
Consequently, the representation of the exercises significantly impacts the model's performance. 
Similarly, removing different intrinsic latent vectors leads to different degrees of performance decline. The finding indicates that processing style is more critical in the initial learning stages while the understanding style are more influential as the learning pattern becomes more fixed.

\end{itemize}

\subsection{RQ3: Sensitivity Analysis of Hyperparameters}
\begin{figure}[htbp]
     \centering
     \resizebox{\linewidth}{!}{
     \begin{minipage}{0.3\textwidth}
        \centering
        \includegraphics[height=3.3cm]{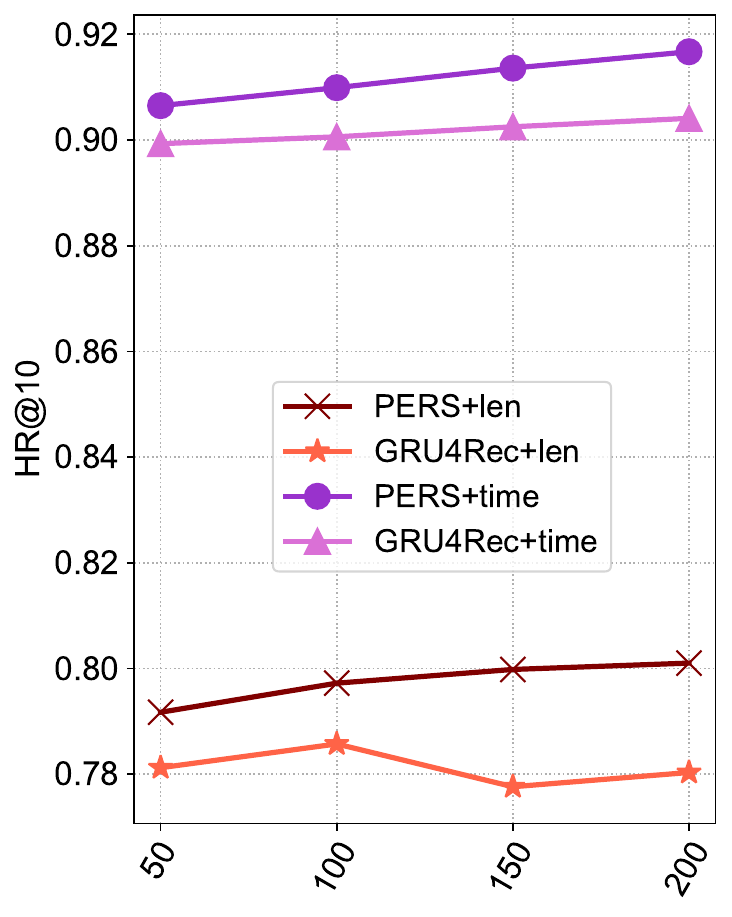}
        \begin{minipage}{\linewidth}
            \centering
              (a) Sequence Length $\lambda$
        \end{minipage}
        \label{fig:chap4_exp_hyp_lambda}
     \end{minipage}
     \hfill
     \begin{minipage}{0.32\textwidth}
        \centering
        \includegraphics[height=3.3cm]{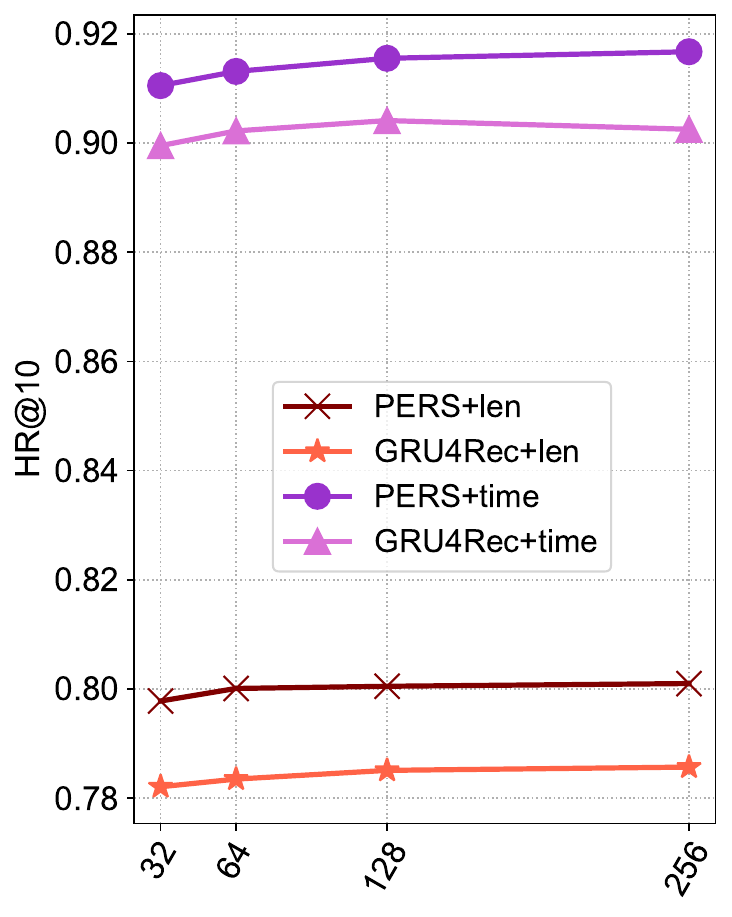}
        \begin{minipage}{\linewidth}
            \centering
              (b) Exercise Embedding $d_p$ 
        \end{minipage}
         \label{fig:chap4_exp_hyp_dp}
     \end{minipage}
     \hfill
     \begin{minipage}{0.3\textwidth}
        \centering
        \includegraphics[height=3.3cm]{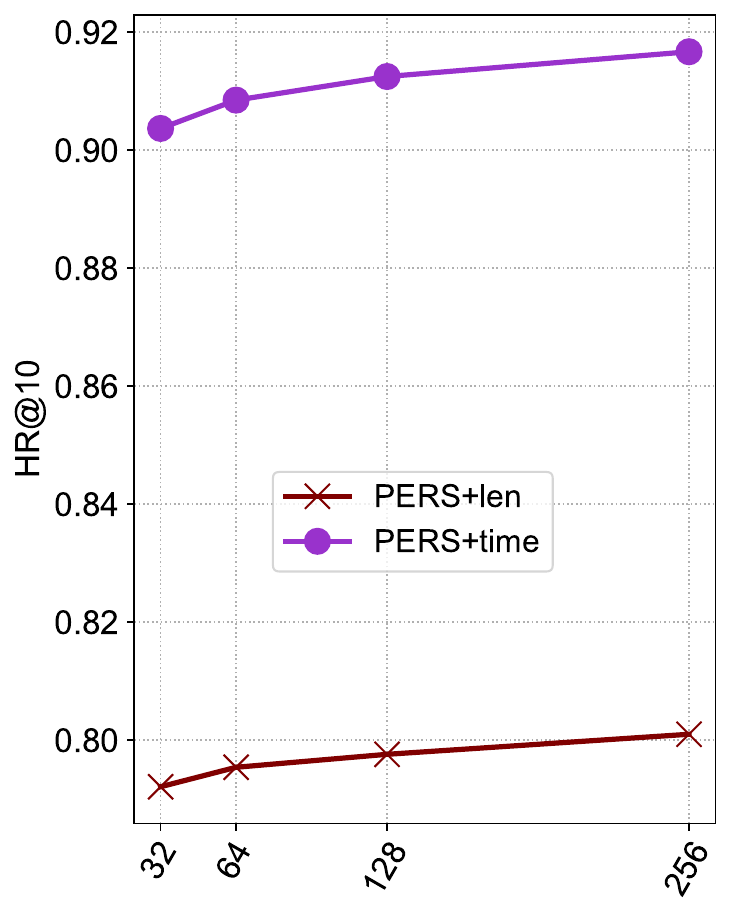}
        \begin{minipage}{\linewidth}
            \centering
              (c) Code Embedding $d_c$
        \end{minipage}
        \label{fig:chap4_exp_hyp_dc}
     \end{minipage}
     }
    \caption{Influence of three key hyperparameters on the performance of the \pers. 
    }
    \label{fig:chap4_exp_hyp}
    
\end{figure}

We conduct a sensitivity analysis on the hyperparameter of PERS with two datasets: CodeNet-len and Code-time. In particular, we study three main hyperparameters: 1) sequence length $\lambda \in \{50, 100, 150, 200\}$, 2) dimension of exercise embedding $d_p\in \{32, 64, 128, 256\}$, and 3) dimension of code embedding $d_c\in \{32, 64, 128, 256\}$. In our experiments, we vary one parameter each time with others fixed. Figure~\ref{fig:chap4_exp_hyp} illustrates the impacts of each hyperparameter and we can obtain the following observations: \par

\begin{itemize}
    \item 
Our model is capable of capturing long sequence dependencies. In Figure~\ref{fig:chap4_exp_hyp}(a), PERS performs better as the sequence length increases, while the results of GRU4Rec remain unchanged or even decline.
 \par
    \item 
    As shown in Figure~\ref{fig:chap4_exp_hyp}(b), the performance of both PERS and GRU4Rec initially improves and then declines as the dimension of exercise embedding increases. The optimal performance is achieved at approximately $d_p=128$. \par
    \item 
As the dimension of code embedding increases, the performance of PERS in Figure~\ref{fig:chap4_exp_hyp}(c) shows a consistent enhancement, highlighting the significance of code features in capturing programming learning patterns.

\end{itemize}

\subsection{RQ4: Case Study on Visualization Analysis}

\begin{figure}
     \centering
     \begin{minipage}{1\textwidth}
        \centering
        \includegraphics[height=2.2cm]{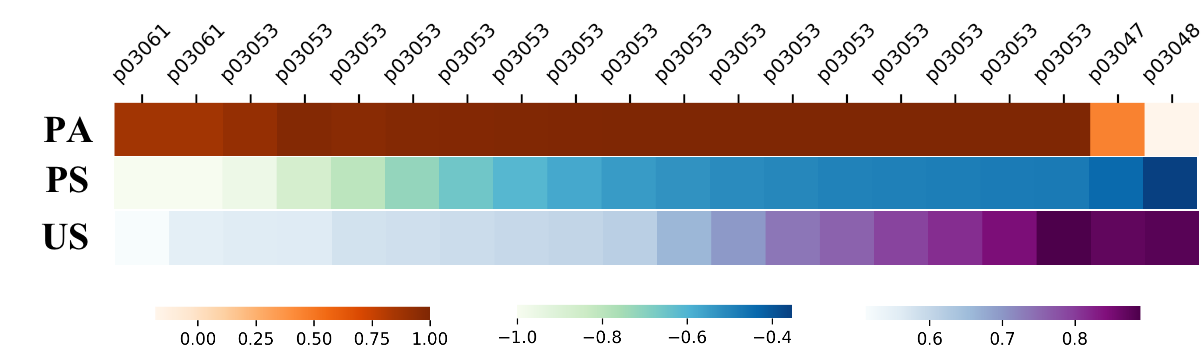}
        \begin{minipage}{\linewidth}
            \centering
              (a) $\mathbf{u}_{222602662}$ 
        \end{minipage}
         \label{fig:chap5-visual-stu1}
     \end{minipage}
     \\
     \begin{minipage}{1\textwidth}
        \centering
        \includegraphics[height=2.2cm]{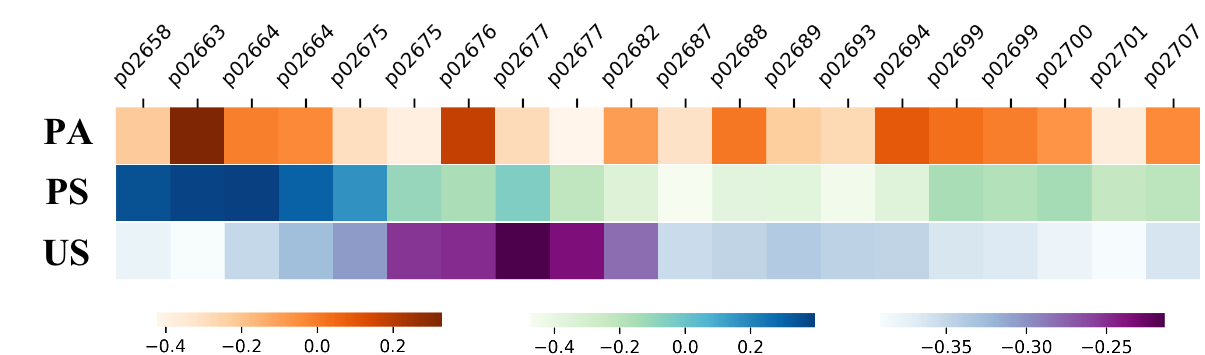}
         \begin{minipage}{\linewidth}
            \centering
              (b) $\mathbf{u}_{737111725}$
        \end{minipage}
         \label{fig:chap5-visual-stu2}
     \end{minipage}
     \caption{Case study on latent vectors visualization}
    \label{fig:chap5-visual}
\end{figure}

%
To demonstrate the interpretability of our approach, we conduct a visualization analysis of three latent vectors involved in the \pers, i.e., programming ability $\mathbf{PA}_{t}$, processing style $\mathbf{PA}_{t}$ and understanding style $\mathbf{US}_{t}$.
We randomly selected the behavioral sequences of two students from the CodeNet dataset for the case study.
From the exercise sequence of each student, we can observe that $\mathbf{u}_{222602662}$ tends to make multiple attempts at the same exercise and solve problems in a systematic manner, while $\mathbf{u}_{737111725}$ prefers solving problems by leaps and bound.
%
We extract these three instinct vectors from the last time step of the model and visualize the dimensional reduction results in Figure\ref{fig:chap5-visual}.
We note some observations in the visualization results:
%

\begin{itemize}
    \item 
The extent of variation in students' programming abilities differs between inter-exercise and intra-exercise. 
Taking $\mathbf{u}_{222602662}$ as an example, as he made multiple attempts on $\mathbf{p}_{03053}$, his programming ability continuously improved. 
However, when he attempted the next exercise, his $\mathbf{PA}_{t}$ showed a noticeable decline.
Therefore, fine-grained modeling of inter-exercise contributes to better capturing students' learning state.
    \item 
The changing patterns of learning styles among different students is consistent with their learning process. 
For $\mathbf{u}_{222602662}$, the value of $\mathbf{PS}_{t}$ and $\mathbf{US}_{t}$ gradually approach 1 during the programming learning process, 
suggesting a reflective and sequential learning style. 
As for $\mathbf{u}_{737111725}$, his corresponding latent vectors exhibit a gradual tendency towards -1, indicating an active and global learning style.
This shows that the latent vectors can learn valuable information, thereby validating the rationality of our model.
\end{itemize}

\section{Conclusions}
In this paper, we study programming exercise recommendation (PER) to enhance engagement on online programming learning platforms.
To solve PER, 
we propose a novel model called \pers\ based on simulating learners' intricate programming behaviors.
First, we extend the Felder-Silverman learning style model to the programming learning domain and present the programming learning style. 
After that, 
based on the programming learning style,
we construct latent vectors to model learner's states, including programming abilty, 
processing style,
and understanding style.
In particular,
we introduce a differentiating module to update the states based on enhanced context, which are positions for exercises and compilation results for codes, respectively.
Finally, 
the updated states at the last time step are sent to predict.
Extensive experiments on two real-world datasets demonstrate the effectiveness and interpretability of our approach.
In future work, we will explore incorporating the difference of structural features from students' submitted code to further enhance the performance of the model.

\subsubsection{Acknowledgements} This work has been supported by the National Natural Science Foundation of China under Grant No.U1911203, and the National Natural Science Foundation of China under Grant No.62377012.

%
%
%

\bibliographystyle{splncs04}
\bibliography{main}

%





\end{document}